\documentclass[12pt]{iopart}

\usepackage{graphicx}

\begin{document}

\title[]{$1/f$ noise on the brink of wet granular melting}

\author{Kai Huang}

\address{Experimentalphysik V, Universit\"at Bayreuth, 95440 Bayreuth, Germany}
\ead{kai.huang@uni-bayreuth.de}
\vspace{10pt}
\begin{indented}
\item[]July 2015
\end{indented}

\begin{abstract}
The collective behavior of a two-dimensional wet granular cluster under horizontal swirling motions is investigated experimentally. Depending on the balance between the energy injection and dissipation, the cluster evolves into various nonequilibrium stationary states with strong internal structure fluctuations with time. Quantitative characterizations of the fluctuations with the bond orientational order parameter $q_{\rm 6}$ reveal power spectra of the form $f^{\alpha}$ with the exponent $\alpha$ closely related to the stationary states of the system. In particular, $1/f$ type of noise with $\alpha\approx-1$ emerges as melting starts from the free surface of the cluster, suggesting the possibility of using $1/f$ noise as an indicator for phase transitions in systems driven far from thermodynamic equilibrium.
\end{abstract}

%
%
%
%
%

\section{Introduction}

$1/f$ noise (also called flicker or pink noise) is a class of signal exhibiting power spectrum $f^{\alpha}$ with exponent $\alpha\sim -1$, typically in the range $-1.4<\alpha<-0.8$~\cite{Weissman1988}. Due to its ubiquity in nature, $1/f$ noise has drawn much more attentions from various communities for almost a century, in comparison to white ($\alpha=0$), Brownian ($\alpha=-2$) as well as diffusion noises~\cite{Press1978,vanVliet1958,Brophy1987,Kiss1997}. The first and most extensively investigated noise of this class is electric noise in conducting or semiconducting materials~\cite{Weissman1988, Schottky1926, Miller1947, Dutta1981}. Later on, the ubiquity of $1/f$ noise is unveiled as a surprisingly large number of systems are found to fall into this class, ranging from blinking of stars and sunspot activity~\cite{Press1978} in astrophysics, earthquake triggering~\cite{Scholz1990} and undersea ocean currents in geophysics~\cite{Schertzer1991}, to the loudness fluctuations of music~\cite{Voss1975}, gene expression~\cite{Furusawa2003}, and human cognition~\cite{Gilden1995,Wagenmakers2004}. 

Because of this ubiquity, it is important to understand why such a common feature persists and to explore whether there is a universal mechanism behind or not. Following the widely used the Bernamont-Surdin-McWhorter (BSM) model~\cite{Bernamont1937,Surdin1939,Ziel1950,McWhorter1957}, $1/f$ noise can be considered as a superposition of independent events with Lorentzian spectra and a broad distribution of relaxation time. Although it was originally introduced to explain electric noises, this model has been successfully applied to a large variety of systems exhibiting $1/f$ noise~\cite{Weissman1988,Jensen1998}. Nevertheless, recent investigations also reveal that the assumption of elementary events with Lorentzian spectra may not be necessary, since models based on random or point process~\cite{Gingl1989,Kaulakys2005} have also been reported to produce $1/f$ noise.

In order to provide a general understanding of $1/f$ noise, Bak \textit{et al.}~proposed the concept of self-organized criticality (SOC)~\cite{Bak1987,Bak1988}, claiming that in spatially extended dissipative systems $1/f$ noise can be viewed as an indication of self-organized critical state. In contrast to the critical state in equilibrium thermodynamics, the system is self-organized, i.e.~no external tuning is needed. Using a cellular automata (CA) model describing the avalanches in a pile of sand, they demonstrated that the scale invariance in time is associated with fractal structures, i.e.~scale invariance in space. Triggered by this concept, a tremendous amount of investigations have been conducted to test the universality of SOC~\cite{Jensen1998,Nagel1992,Bak1999}. However, the CA model proposed by BTW was soon found to produce $1/f^2$ noise in both one- and two-dimensions~\cite{Jensen1989,Kertesz1990} and thus cannot directly be used to explain $1/f$ noise. Moreover, contradictory results were found in model experiments on the avalanches of a sandpile~\cite{Nagel1992,Held1990,Jaeger1989}. Those investigations led to the conclusion that SOC is not as universal as it was claimed. Instead, its validity relies on the dissipation mechanisms of specific systems~\cite{Frette1996}, and also on the way of analysis~\cite{Buchholtz1994,Buchholtz1996}. Thus from the SOC perspective, the ubiquitous $1/f$ noise is still not completely understood and whether a general theory for $1/f$ noise exists or not is still unclear. Nevertheless, the advance on the relation between the invariance of time and space gives us the hint that \emph{noise} may provide useful insights into the self-organized stationary states of systems driven far from thermodynamic equilibrium~\cite{Jensen1998}. 

Wet granular matter, such as wet sand used to build sand sculptures, has been drawing more and more interest from both physical and engineering communities in the past decades~\cite{Iveson2001,Mitarai2006,Herminghaus2013}. This is not only because of its ubiquity in nature, industries and our daily lives, but also due to the fact that it can be treated as a nonequilibrium model system  with granular particles replacing molecules and liquid bridges formed between adjacent particles replacing molecular bonds. Because of the strong energy dissipation associated with wet impacts~\cite{Gollwitzer2012}, continuous energy injection is necessary to maintain a certain stationary state of such a nonequilibrium system. Former investigations have revealed that the dissipative energy scale arising from the rupture of liquid bridges between adjacent particles plays an important role in determining the collective behavior of wet granular matter, such as melting~\cite{Scheel2004,Huang2009a}, clustering~\cite{Huang2012}, and phase separation~\cite{Fingerle2008,Huang2009b}. More recently, an analogue of surface melting was found in agitated wet granular matter~\cite{Roeller2011,May2013}, suggesting that existing knowledge on phase transitions in equilibrium systems can be extended to explain the wide-spreading nonequilibrium systems in nature. Here, focusing on the noise spectra extracted the nonequilibrium stationary states of a wet granular cluster, I demonstrate the possibility of using $1/f$ type noise as an indicator for phase transitions far from thermodynamic equilibrium.

\section{Methods}

\begin{figure}
  \centering
  \includegraphics[width = 0.65\textwidth]{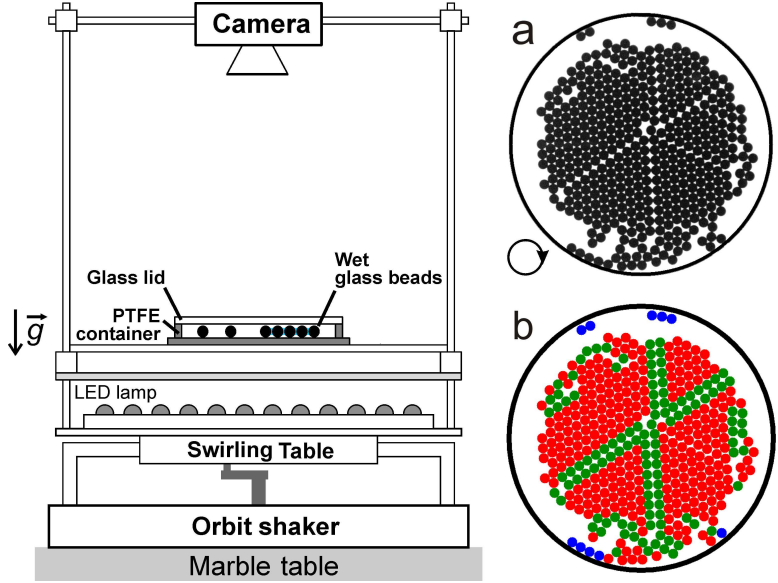}
  \caption{A sketch of the experimental setup. The swirling table provides circular motion of the whole container in a plane perpendicular to the gravitational acceleration $\vec{g}$. Both the CCD camera and the LED lamp are mounted in the co-moving frame of the swirling table. (a) A typical snapshot of a wet granular ``crystal'' with defects. The circular arrow denotes the swirling motion of the container. (b) The processed image with particles color coded according to their local structures: Red, green and blue represent hexagonal, square and chain structures, respectively.}
  \label{fig:setup}
\end{figure}

The experimental setup is sketched in Fig.~\ref{fig:setup}. Various number $N$ of polished soda lime glass beads (SiLibeads P) with a density of $\rho_{\rm p}=2.58~{\rm g/cm^{3}}$ and a diameter of $d=4\pm0.02$~mm are used as granular sample. After being mixed with a certain volume $V_{\rm l}$ of purified water (specific resistance $18.2$\,M$\Omega$cm, \mbox{LaborStar TWF}), the particles are added into a cylindrical polytetrafluoroethylene (PTFE) container with an inner diameter of $D=102$~mm and a height of $6$~mm. The latter confinement ensures a strict mono-layer of particles. The liquid content $W=V_{\rm l}/V_{\rm s}$ with $V_{\rm s}$ the total volume of the spheres is fixed at $0.02$, which ensures short ranged attractive interactions between adjacent particles via the formation of liquid bridges~\cite{Willet00}. The container is closed with a glass lid to keep the liquid content constant. During the experiments, the glass lid is slightly heated to avoid water vapor condensed there. The wet granular sample is illuminated from below with a LED array, and viewed from top with a CCD camera (Lumenera LU125M) mounted in the comoving frame [see Fig.~\ref{fig:setup}(a) for a typical snapshot]. The swirling table is leveled within $5.7\times10^{-3}$ degrees to avoid the influence from gravity. The horizontal swirling motion, i.e.~a superposition of two perpendicular sinusoidal vibrations with identical amplitude $A$ and a phase difference of $\pi/2$, is provided by a mechanical orbit shaker (Thermolyne, AROS160). The amount of energy injection into the system is adjusted by varying the driving frequency $f_{\rm d}$ at a fixed amplitude $A=31.8$~mm. With a computer controlled precision resistance decay box (Burster 1424), the swirling frequency is controlled with an accuracy of $\sim10^{-4}$~Hz. $f_{\rm d}$ is measured by tracing a fixed particle on the swirling table with another CCD camera (Lumenera LU075M) mounted in the lab frame. A computer program is developed to adjust the swirling frequency in steps and capture images with a fixed rate of $5$ frames per second. 

Using an image processing procedure based on the Hough transformation~\cite{Kimme75}, I determine the positions of all particles in each frame captured. For each particle, the bond orientational order parameters (BOOP) are calculated with~\cite{Steinhardt83,Wang05,Lechner2008} 

\begin{equation}
\label{bop}
q_{\rm l}=\sqrt{\frac{4\pi}{2l+1}\sum_{m=-l}^{l}|\bar{Q}_{\rm lm}|^2},
\end{equation}

\noindent where $\bar{Q}_{\rm lm}\equiv \langle Q_{\rm lm}(\vec{r})\rangle$ is an average of the local order parameter $Q_{\rm lm}(\vec{r})\equiv Y_{\rm lm}(\theta(\vec{r}),\phi(\vec{r}))$ 
over all bonds connecting this particle to its nearest neighbors, which are identified with a critical bond length $r_{\rm b}=1.35\,d$. Here $Y_{\rm lm}(\theta(\vec{r}),\phi(\vec{r}))$ corresponds to the spherical harmonics of a bond located at $\vec{r}$. 
Finally, I determine the local structure of each particle by comparing the order parameter $q_{\rm 6}$ with the standard values for perfectly hexagonal, square, line structures, as well as for isolated particles. Here, $q_{\rm 6}$ is chosen as the order parameter because of its sensitivity to the hexagonal order. As shown in Fig.~\ref{fig:setup}(b), the particles are colored by their local structures after the above analysis in order to visualize the local structure changes.

The collective behavior of a wet granular cluster is obtained through initializing the system at a relatively large $f_{\rm d}=1.745$~Hz for $2$ hours followed by decreasing $f_{\rm d}$ with a step of $\Delta f_{\rm d}=0.062$~Hz and fixed waiting time $\Delta t=700$~s. The long initial time ensures a homogeneous wetting liquid distribution and a stable temperature gradient to avoid the condensation of water on the lid.  Immediately after the lowest driving frequency is reached, $f_{\rm d}$ is swept up with the same $\Delta f_{\rm d}$ and $\Delta t$. The whole cycle is repeated seven times for each particle number $N$.

\section{Nonequilibrium stationary states}

\begin{figure}
  \centering
  \includegraphics[width = 0.8\textwidth]{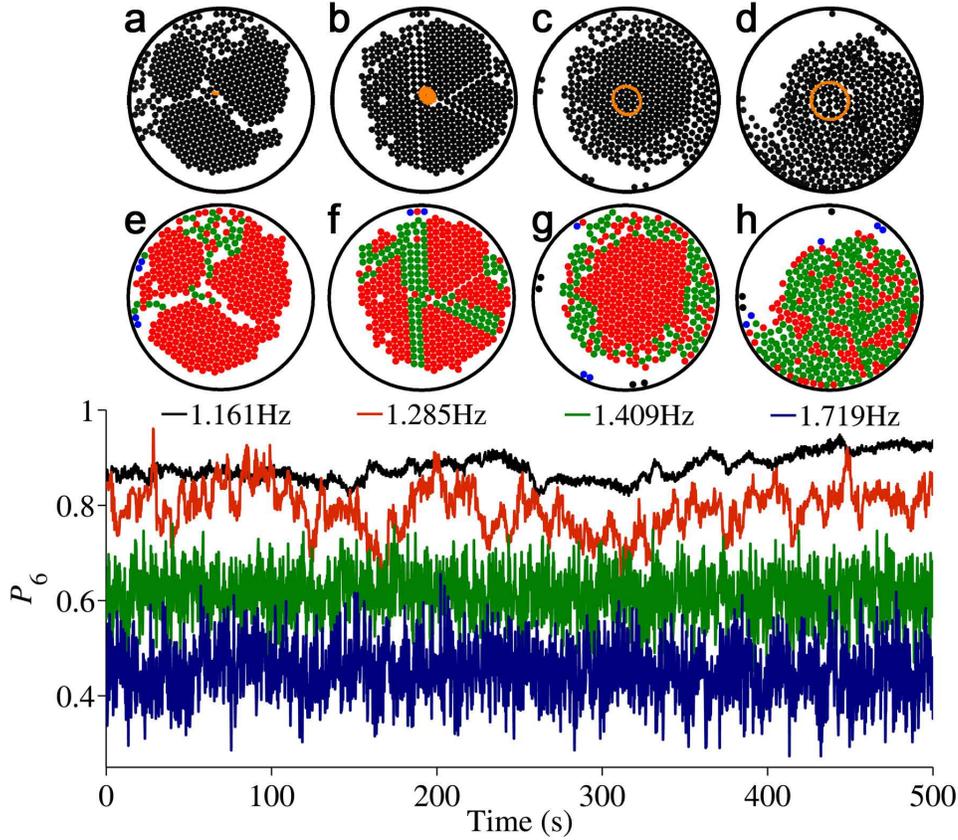}
  \caption{Sample images of a wet granular cluster in various nonequilibrium stationary states and corresponding internal structure fluctuations with time. Upper panels: Snapshots of the cluster in its initial crystalline (a), amorphous (b), surface melting (c), and completely melted (d) states with the positions of its center of mass (c.m.) in the co-moving frame over the whole observation period overlaid as a cloud of orange dots. Middle panels: Corresponding internal structures of the cluster based on an analysis with BOOP. The color code is the same as described in the caption of Fig.~\ref{fig:setup}. Isolated particles are colored in black. Lower panel: The fluctuations of the order parameter $P_{\rm 6}$, i.e.~fraction of particles in the hexagonal structure (in red), with time. The total number of particles $N=406$, corresponding to an area fraction $\Phi=0.624$.}
\label{fig:evolution}
\end{figure}

Figure~\ref{fig:evolution} shows the collective behavior of a wet granular cluster at four different driving frequencies and corresponding internal structure fluctuations. The internal structure is characterized by $P_{\rm 6}$, the percentage of particles in a local hexagonal structure. 

At $f_{\rm d}=1.161$~Hz, the cluster is composed of a few ``crystals'' loosely connected with each other. Here, ``crystals'' refer to clusters composed of particles in a local hexagonal packing. As the positions of the cluster center [see the cloud of dots in Fig.~\ref{fig:evolution}(a)] indicate, the mobility of the cluster is low. This suggests that the external driving force is not sufficient for the ``crystals'' to overcome the friction from the ground and move freely. The $P_{\rm 6}$ order parameter stays at a high value close to $0.9$, indicating that almost all particles are kept in a hexagonal structure. Thus, this state is called initial crystalline state. The small fluctuations of $P_{\rm 6}$ mainly arise from the interactions between the loosely connected ``crystals'' and the frequent attaching and detaching of individual particles from them.

As the driving frequency $f_{\rm d}$ increases to $1.285$~Hz, the enhanced mobility of the particles leads to the merging of the ``crystals'' into a single one with dislocation defects. The defect lines, along which local structures of particles deviate from hexagons to squares, suggest a certain amount of potential energy being stored there. This energy is obtained from the frequent collisions between the ``crystal'' and the container wall, because of the higher mobility of the whole cluster in comparison to the initial crystalline state. Note that the state with defects is unstable: A small perturbation will lead to a healing of the defects due to the cohesion between adjacent particles. As the ground state with a perfect hexagonal packing is not easily achieved, the healing process typically ends up with a metastable state corresponding to a local potential energy minimum. The corresponding fluctuations of $P_{\rm 6}$ are much stronger than the initially crystalline state, because the defects associated structure changes occur at the length scale of the whole ``crystal''. Moreover, the fluctuations occur at time scales much longer than the swirling period, i.e.~$1/f_{\rm d}$. 


Such type of fluctuations persist as the driving frequency increases, as long as the cluster stays in the amorphous state. As $f_{\rm d}$ increases further to $1.409$~Hz, an abrupt transition into the surface melting state arises, owing to a balance between the energy injection and the rupture energy of a liquid bridge~\cite{May2013}. This transition is manifested by the collapse of the voids inside the ``crystal'' together with the emerging of a liquid like layer covering the crystalline core, as shown in Fig.~\ref{fig:evolution}(c) and (g). Consequently, the whole cluster moves in a circular trajectory with permanent contact to the container wall (Note that the shortest distance from the circle to to the container wall matches the mean radius of the cluster). This feature indicates that the hard impacts between the cluster and the container wall with a finite effective coefficient of restitution (COR) is now replaced with soft ones (i.e.~zero COR), because the liquid like layer acts as an effective damper.

It is remarkable to see that the fluctuations of $P_{\rm 6}$ also behave dramatically different in the surface melting state in comparison to the amorphous one: The large fluctuations at long time scales in the amorphous state are not present any more. This can be understood from the fact that the energy injection and dissipation in this state is localized in the molten surface layer of the cluster, because of the much more frequent collisions there. Hence, the metastability associated with the change of the crystalline structures at a length scale of the whole cluster is suppressed, leading to a white noise type fluctuations. Further increasing $f_{\rm d}$ leads to an growth of the molten layer thickness, until the whole cluster  behaves like a liquid at $f_{\rm d}=1.719$~Hz. Along with the growth of the molten layer thickness, the center of the more deformable cluster is driven closer to the rim of the container, as a comparison between Fig.~\ref{fig:evolution}(c) and (d) reveals. Moreover, the transition into the liquidlike state leads to a decrease of the average $P_{\rm 6}$, although the behavior of the fluctuations persists.

The various stationary states and the associated structure fluctuations are robust: Variations of $\Delta f_{\rm d}$, $\Delta t$ by at least one order of magnitude and a change the sweeping direction of $f_{\rm d}$ yield the same behavior. The lower limit of $f_{\rm d}$ is selected such that the wet granular assemblies are immobile, i.e.~the driving force cannot overcome the total frictional force from the ground.

\section{Power spectrum of internal structure fluctuations}

\begin{figure}
  \centering
  \includegraphics[width = 0.75\textwidth]{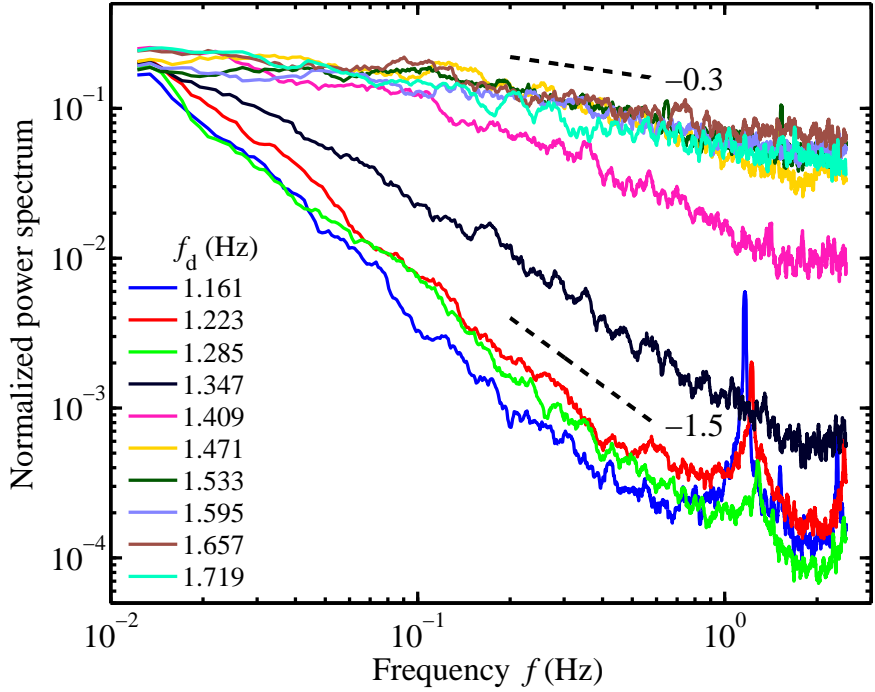}
   \caption{Normalized power spectrum of $P_{\rm 6}$ at various driving frequency $f_{\rm d}$. The measurement is performed by decreasing $f_{\rm d}$. Each power spectrum is an average over $7$ experimental runs. To enhance visibility, the data are smoothed with running average. The dash lines (corresponding to power law exponent $-0.3$ and $-1.5$) are drawn to guide the eyes. Other parameters are the same as described in the caption of Fig.~\ref{fig:evolution}.}
  \label{fig:spec}
\end{figure}

From the perspective of $1/f$ noise, it is intuitive to analyze the power spectrum of the internal structure fluctuations in various nonequilibrium stationary states of the wet granular cluster. As shown in Fig.~\ref{fig:spec}, the power spectra of $P_{\rm 6}$ exhibit predominately power law behavior for a wide range of $f_{\rm d}$. In the initial crystalline or amorphous state ($f_{\rm d} \leq 1.285$~Hz), all the power spectra decay with an exponent of $\alpha\approx-1.5$. The peak at $f=f_{\rm d}$ indicates the influence from the driving frequency, as the energy injection is associated with periodic collisions between the cluster and the container wall. The peak becomes less and less pronounced as $f_{\rm d}$ grows, which can be attributed to the ``softening'' of the impact: As the ``crystal'' becomes more amorphous, the effective COR for its impact with the container wall decreases, and the internal structure change of the cluster becomes more susceptible. This trend will eventually lead to more random collisions with the container wall and consequently a vanishing influence from the container wall. As $\alpha=-1.5$ is typically associated with noise generated by diffusion mechanisms~\cite{vanVliet1958,Brophy1987,Kiss1997}, it is interesting to explore possible diffusion mechanisms of driven wet granular particles in an amorphous state. For a monolayer of noncohesive dry granular particles, former investigations (see e.g. Reis \textit{et al.}~\cite{Reis2007}) reveal a non-trivial subdiffusive behavior in the vicinity of melting transition, which arises from the caging dynamics. However, it is still unclear what the diffusion mechanisms of wet granular particles in various nonequilibrium stationary states are and how it is related to the power spectra observed here. Further clarifications of these questions require a detailed characterization of the particle mobility and will be a focus of further investigations.

The exponent $\alpha$ decreases to $\approx-1.2$ as $f_{\rm d}$ increases to $1.347$~Hz, suggesting a $1/f$ type noise in the vicinity of melting transition. In the surface melting regime with even higher $f_{\rm d}$, $\alpha$ grows continuously until it saturates at $\approx-0.3$, suggesting a trend toward a white noise type of fluctuations without any long time correlations. Meanwhile, a cut-off frequency of $\approx0.1$~Hz, below which the power spectrum decays much slower, arises. 

\begin{figure}
  \centering
  \includegraphics[width = 0.6\textwidth]{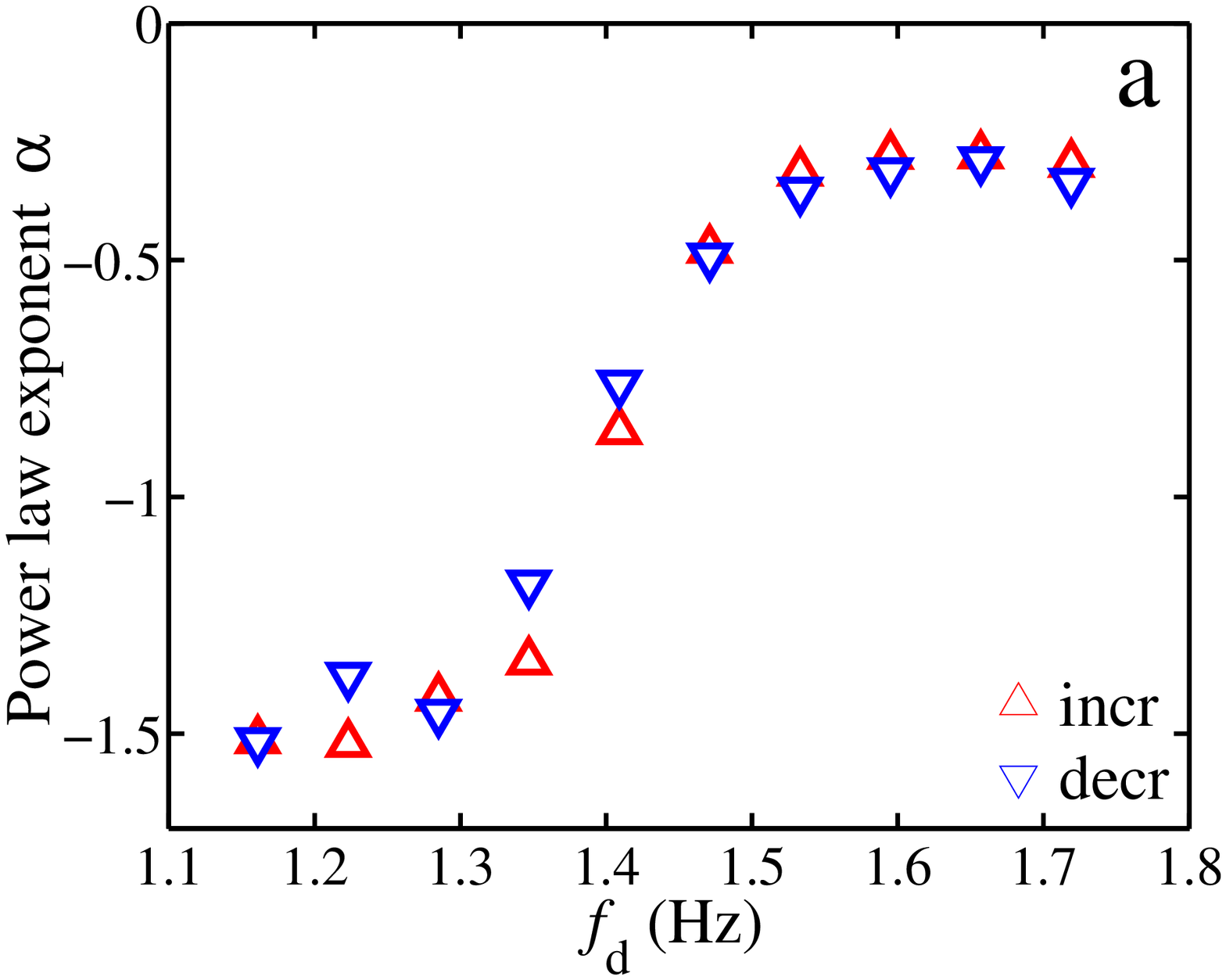} \\
  \includegraphics[width = 0.6\textwidth]{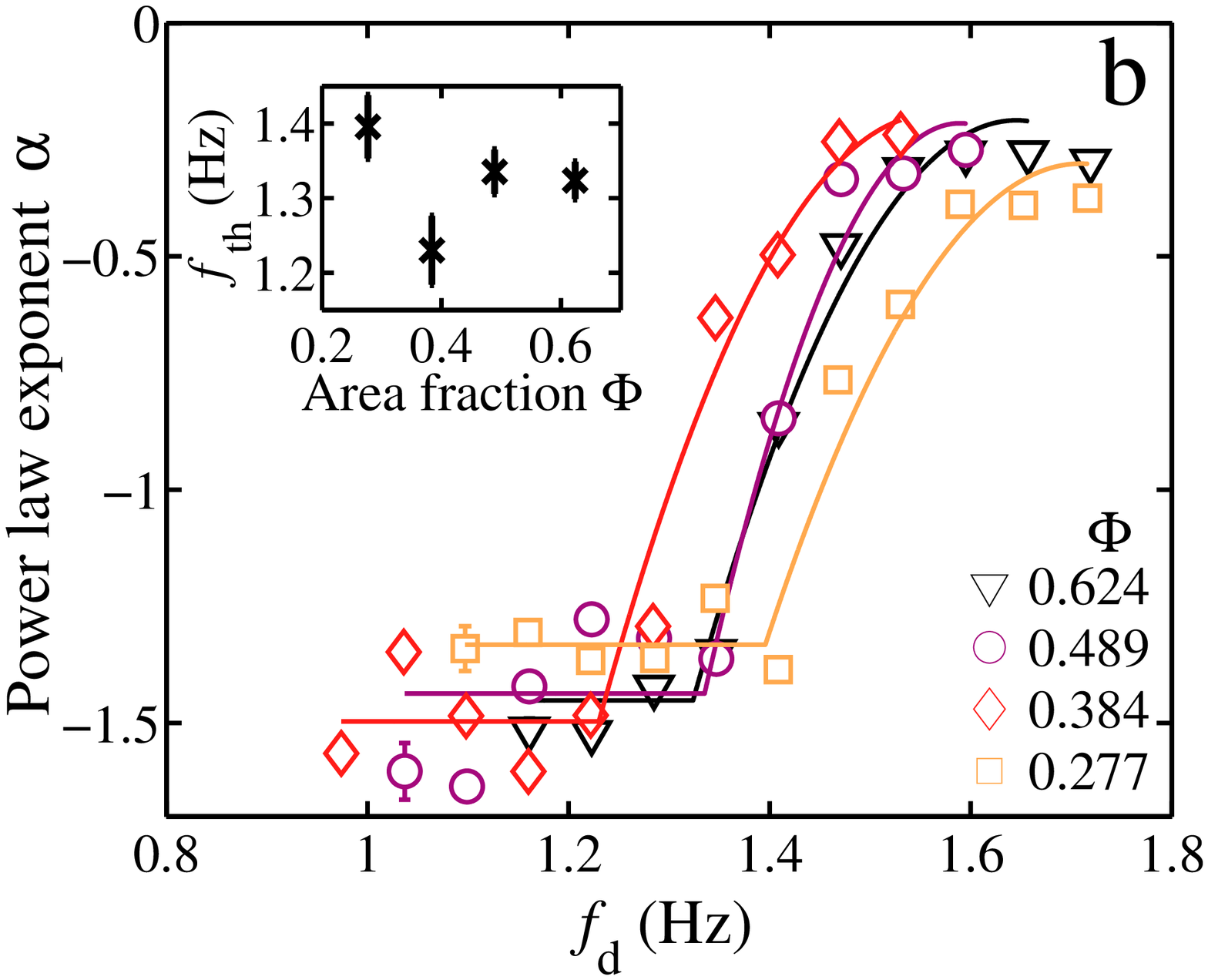}
  \caption{(a) Power law exponent $\alpha$ as a function of the driving frequency $f_{\rm d}$ for area fraction $\Phi=0.624$ ($N=406$). Upper and lower triangles represent data obtained by increasing and decreasing $f_{\rm d}$, respectively. (b) $\alpha$ as a function of $f_{\rm d}$ for various $\Phi$. Solid lines are fits to the data for determining the threshold frequency $f_{\rm th}$ (see text for details). Inset shows the threshold frequency $f_{\rm th}$ as a function of $\Phi$.} 
  \label{fig:scaling}
\end{figure}

As the power law exponent $\alpha$ is closely related to the stationary states of the system, it is intuitive to use it as an order parameter for the melting transition. Figure~\ref{fig:scaling}(a) (lower triangle) shows the fitted power law exponent as a function of the driving frequency for the spectrum shown in Fig.~\ref{fig:spec}. The transition from an amorphous state to a melted state can be clearly distinguished from the deviation of $\alpha$ from its initial low value. To avoid the influence from the peak at $f_{\rm d}$, I fit the spectrum with an upper limit $f_{\rm lim}$, which is determined by decreasing $f_{\rm lim}$ in steps from $f_{\rm d}$ until the standard error of the fit converges. Because of the logarithmic scale of the power spectra, the data below the cut-off frequency play a minor role in determining $\alpha$. Thus,a lower limit of $f$ is not included in the fitting algorithm. A comparison to the results obtained by increasing $f_{\rm d}$ (upper triangle) shows a good agreement, except for the region where the exponent starts to deviate from its initial low value. In this region, the slightly smaller $\alpha$ for the case of increasing $f_{\rm d}$ indicates hysteresis, which is in agreement with a former investigation~\cite{May2013}.

To check possible influence of the finite system size, I vary the particle number from $180$ to $406$, which corresponds to a range of area fraction from $0.277$ to $0.624$. As shown in Fig.~\ref{fig:scaling}(b), similar behavior of the power law exponent is found for all the area fractions explored. This demonstrates the robustness of using $1/f$ noise as an order parameter for the melting transition. More specifically, the trend of a rapid deviation from the initially low $\alpha\approx-1.5$ into a saturated region persists. Hence, I fit the data with a constant value followed by a parabolic growth, and determine the threshold $f_{\rm th}$ as the intersection point that minimizes the standard error. Quantitatively, the threshold frequency stays at $\approx1.3$~Hz, with larger error bars for small cluster sizes. Note that the smallest cluster with $N=180$ corresponds to a radius of roughly $7$ particle diameters. For such a small ``crystal'', the energy obtained from the container wall presumably leads to relatively strong fluctuations of the internal structures, and consequently large data scattering for the initial crystalline and amorphous states. For even smaller cluster size, the reduced probability for all particles to stay in one cluster hinders the statistical measurements performed here.

In the following part of this section, I try to provide a clue to the emerging $1/f$ noise and its connections to the dynamics of the wet granular clusters in the vicinity of the melting transition. As a starting point, we need to identify the essential ingredients leading to the $1/f$ noise. As already discussed in the above analysis, each assembly of the wet granular particles can be considered as one metastable state. This is reminiscent to the packing of equal sized spherical particles in three dimensions: The face-centered cube packing (fcc) with a density of $0.74$ is rare~\cite{Hales2005}. From time to time, we obtain a distribution of metastable states with a packing density range from random close to loose packings~\cite{Scott1960,Nowak1998}. 

The impacts with the container wall lead to frequent switches of the cluster from one metastable state to another one emerge, which can be considered as individual activation-relaxation processes. For example, the collision of a cluster in the amorphous state with the container wall leads to a certain energy injection $E_{\rm inj}\propto(1-\epsilon^2)$, where $\epsilon$ is the effective COR. On one hand, $E_{\rm inj}$ leads to an activation of defects inside the cluster, a corresponding reduction of $P_{6}$, and an increase of the total potential energy through breaking of liquid bridges. On the other hand, it effectively ``heats'' the cluster up, i.e.~enhances the kinetic energy of granular particles therein. The latter effect leads to a healing of defects generated, because the structure with four fold rotational symmetry along the dislocation defects is locally unstable (see Fig.~\ref{fig:evolution} and corresponding discussions). Moreover, the additional kinetic energy obtained from the healing process of a large defect may in turn generate small defects. A cascade of such energy transfer processes toward smaller and smaller length scales eventually leads to a relaxation of the injected energy through the dissipative interactions between individual particles. Following the BSM model~\cite{Bernamont1937,Surdin1939,Ziel1950,McWhorter1957}, we may assume that the relaxation process after an impact has the following autocorrelation function

\begin{equation}
C(\tau)=A\cdot e^{-\frac{\tau}{\tau_0}},
\end{equation}

\noindent where $A$ and $\tau_0$ denote the activation energy and the relaxation time, respectively. Note that the two parameters may differ dramatically from one impact to another, because the relaxed state after one impact is not necessarily the one with the lowest potential energy. Instead, a distribution of various metastable states arises with each state corresponding to a certain configuration of the wet granular particles. 

As $C(\tau)$ is an even function, one can then calculate the corresponding power spectrum for this process with
 
\begin{equation}
P_{\rm i}(f)=2\int_0^\infty A e^{-\frac{\tau}{\tau_0}} \cos(2\pi f \tau){\rm d}\tau,
\end{equation}

\noindent which yields the Lorentzian function

\begin{equation}
P_{\rm i}(f)=\frac{2A \tau_0}{1+(2\pi f \tau_0)^2}.
\end{equation}

\noindent Assuming individual impacts are independent with each other, the power spectrum $P(f)$ for the structure fluctuations measured can be obtained as a superposition of Lorentzian processes 

\begin{equation}
P(f)=\int_0^\infty \frac{2A \tau_0}{1+(2\pi f \tau_0)^2}\cdot {\rm Pr}(A, \tau_0) {\rm d} \tau_0,
\end{equation}

\noindent where the probability ${\rm Pr}(A, \tau_0)$ of having activation energy $A$ and relaxation time $\tau_{\rm 0}$ crucially determines the exponent of $P(f)$. For example, the distribution ${\rm Pr}(A, \tau_0)\propto 1/\tau_0$ leads to $P(f)\propto 1/f$ for the whole frequency range. In reality, a limit on the relaxation time, say $\tau_0\in[\tau_1,\tau_2]$, will lead to a limit on the power law decay: The spectrum flattens below $1/\tau_2$ and steepens above $1/\tau_1$. Therefore, one can determine the intrinsic time scale of the relaxation process from the power spectra.

Note that the relaxation process through a cascade of defects nucleation and healing events relies on a certain rigidity of the cluster. As melting starts at relatively large $f_{\rm d}$, the spatial as well the associated time correlation vanishes, and the energy injection and dissipation occur locally in the molten layers close to the impact point. As a consequence, a white noise type fluctuations with vanishing time correlation arise. As the internal structure fluctuations can be considered as a superposition of individual activation-relaxation processes, the dependency of the power law exponent on the stationary states of the nonequilibrium system represents the change of the corresponding correlation time.

So far, we have a qualitative understanding on why the power law exponent can be used to determine the transitions between various stationary states in the nonequilibrium model system. However, one has to face the following obstacles toward a quantitative understanding: There exists possible coupling between individual activation-relaxation events. A hard impact with the container wall will change the structure of a cluster substantially, which in turn will influence the COR for the next impact. Hence the assumption of independent subsequent events may not be justified. Moreover, such an influence will also lead to a certain distribution of the activation factor $A$, i.e.~the energy injection, and a coupling between $A$ and the relaxation time $\tau_{\rm 0}$. Therefore it is necessary to find the relation between $A$ and $\tau_{\rm 0}$ for more quantitative comparisons to the experiments. 

\section{Conclusions}

In conclusion, I investigate the self-organization of a two-dimensional wet granular cluster under horizontal swirling motion, and use the percentage of particles in a local hexagonal packing to characterize the internal structure fluctuations with time. In the frequency domain, the power spectra of all fluctuations exhibit predominately a power law behavior with decay exponent ranging from $\alpha\approx-1.5$ to $\approx-0.3$. More specifically, the exponent is closely related to the nonequilibrium stationary states of the system: For the initial crystalline and amorphous states at low driving frequencies, it stays low at $\approx-1.5$, suggesting a diffusion noise. For the molten state at large driving frequencies, it saturates at $\approx-0.3$, suggesting a trend toward white noise. In the intermediate states associated with the melting transition, $1/f$ type of noise with $\alpha\approx-1$ arises. A variation of the cluster size yields qualitatively the same power law behavior and quantitatively the same dependency of the decay exponent with $f_{\rm d}$. Such a robust behavior suggests an alternative way of detecting phase transitions in systems driven far from thermodynamic equilibrium. Finally, I show that the $1/f$ noise can be qualitatively understood from the impact mechanisms between the wet granular cluster and the container wall. 


This investigation demonstrates the importance of \emph{noise} in characterizing systems driven out of thermodynamic equilibrium. Concerning the robustness of such a characterization, future questions to ask are: Will any type of fluctuation related to the energy flux in and out of the nonequilibrium system, e.g.~the sound energy arising from the inelastic collisions, exhibits a similar behavior? Will the way of energy injection or dimensionality of the system influence the behavior? Concerning a more quantitative understanding of the emerging $1/f$ noise, it is also important to characterize the coupling between the activation energy and the relaxation time experimentally through a measurement of the effective coefficient of restitution for wet granular clusters.   
 
\section{Acknowledgement}

I am grateful to Ingo Rehberg for helpful hints and a critical reading of the manuscript. I thank Michael Wild and Christopher May for their preliminary work on the experimental setup. Inspiring discussions with Victor Steinberg and Mario Liu are gratefully acknowledged. This work is supported by the Deutsche Forschungsgemeinschaft through Grant No.~HU1939/2-1.

\section{References}

 \bibliographystyle{iopart-num}
\providecommand{\newblock}{}

\end{document}